\def\fj     { f_J (2220) }
\def\fpm    { f'_2 (1525) }
\def\gam    { \gamma \gamma }
\def\ksks   { K_s K_s }

\def\pppm   { \pi^+ \pi^- }

\def\ggra   { \gamma \gamma \rightarrow }
\def\ggfj   { \ggra \fj }

\def\ggfj   { \ggra \fj }

\def\Ggg    { \Gamma _{ \gamma \gamma } }

\def\lumi   { 3.0 \, {\rm fb}^{-1} }

\def\qqb    { q \overline{q} }

\def\ggfkk {\gamma\gamma \rightarrow f_J(2220) \rightarrow K_{s,1} K_{s,2}}
\def\ppone   {(\pi^+\pi^-)_1}
\def\pptwo   {(\pi^+\pi^-)_2}

\def\brkkggfj { (\Ggg \, {\cal B}_{\ksks})_{\fj} }
\def\brkkggfpm { (\Ggg \, {\cal B}_{\ksks})_{\fpm} }

\def\brksks { {\cal B}_{\ksks} }

\def\brmcksks { {\cal B}^{MC}_{\ksks} }
\def\brdksks { {\cal B}^{data}_{\ksks} }

\def\setfigsize    { \epsfysize=8cm }
\def\setbigfigsize { \epsfxsize=5in }

\def\setfigsize    { \epsfysize=8cm }
\def\setbigfigsize { \epsfxsize=5in }

\def\brffjkk { {\cal B} (\fj \rightarrow K_sK_s) }
\def\gammajf { \Gamma(J/\psi\rightarrow\gamma \fj)\, \brffjkk}

\def\glim    {\Gamma_{lim}}

\def\footbesres {
We average the mass and width measurements for the
four different modes reported by BES and the two modes
reported by Mark III.  
We assume that the systematic uncertainties within an experiment
are completely correlated and the systematic uncertainties between
experiments are uncorrelated.}

\def\footbespsiwidth {
We form an average $K_s K_s$
branching fraction by combining
the $K_s K_s$ with the $K^+K^-$ measurements. 
We treat systematic uncertainties as 100\% correlated
within an experiment and uncorrelated between
experiments. }

\newif\ifclns \clnsfalse
\newif\ifchklen \chklenfalse

\clnstrue
\chklenfalse

\ifchklen
\def\setfigsize{ \epsfxsize=3.375in}
\else
\def\setfigsize{ \epsfysize=10cm }
\fi

\ifclns
   \ifchklen
      \documentstyle[aps,prl,epsf,floats]{revtex}
   \else
      \chklenfalse
      \documentstyle[aps,prl,preprint,floats,epsf]{revtex}
   \fi
\else
   \documentstyle[aps,prl,epsf,floats]{revtex}
\fi

\begin{document}

\ifclns
   \draft
\fi
%
%

\ifclns
\preprint{\tighten\vbox{\hbox{CLNS 97/1467 \hfill}
                        \hbox{CLEO 97-3   \hfill}
} }
\fi
\title{Limit on the Two-Photon Production of the \\
Glueball Candidate $\fj$ at CLEO}

\author{CLEO Collaboration}

\date{March 15, 1997}
\maketitle


\widetext
\begin{abstract}
\ifclns
   \tighten
\fi

We use the CLEO detector at the Cornell $e^+e^-$ storage ring, CESR,
to search for the two-photon production of the glueball candidate $\fj$ 
in its decay to $\ksks$.  
We present a restrictive upper limit on 
the product of the two-photon partial width
and the $\ksks$ branching
fraction, $\brkkggfj$.
We use this limit to calculate a lower
limit on the stickiness, which is a measure of the
two-gluon coupling relative to the two-photon
coupling.  
This limit on stickiness indicates 
that the $\fj$ has substantial glueball content.

\end{abstract}

\pacs{12.39.Mk,13.65.+i,12.38.Gc }
\narrowtext

\tighten

\clearpage

\ifclns
\begin{center}
R.~Godang,$^{1}$ K.~Kinoshita,$^{1}$ I.~C.~Lai,$^{1}$
P.~Pomianowski,$^{1}$ S.~Schrenk,$^{1}$
G.~Bonvicini,$^{2}$ D.~Cinabro,$^{2}$ R.~Greene,$^{2}$
L.~P.~Perera,$^{2}$ G.~J.~Zhou,$^{2}$
% ...
\end{center}
 
\small
\begin{center}
$^{1}${Virginia Polytechnic Institute and State University,
Blacksburg, Virginia 24061}\\
$^{2}${Wayne State University, Detroit, Michigan 48202}\\
% ...
$^{12}${Carleton University, Ottawa, Ontario, Canada K1S 5B6 \\
and the Institute of Particle Physics, Canada}\\
$^{13}${McGill University, Montr\'eal, Qu\'ebec, Canada H3A 2T8 \\
and the Institute of Particle Physics, Canada}\\
% ...
\end{center}

   \thispagestyle{empty}
   \clearpage
\fi
%
%

        The two-photon width of a resonance is a probe of the
electric charge of its constituents, 
so the magnitude of the
two-photon coupling can serve to distinguish
quark-dominated resonances from glue-dominated resonances 
(henceforth simply called ``glueballs''). 
The $\fj$, 
sometimes referred to as the $\xi(2230)$, 
was first reported by the Mark III collaboration\cite{mark}.
This resonance is a glueball candidate due
to its narrow width \cite{mark,bes},
its observation in glue-rich environments \cite{mark,bes,gams,lass,mss}, 
and its proximity in mass to lattice QCD predictions 
of the tensor glueball \cite{michael,morningstar}.

In this Letter we report on a search for the $\fj$ in two-photon 
interactions at CLEO
and set an upper limit on
the product of 
its two-photon partial width and
branching fraction to $\ksks$,
improving on a previous limit
set by ARGUS \cite{argus} using the $K^+ K^-$ decay mode.
Using our measurement, 
we calculate the stickiness,
a useful glueball figure of merit \cite{chanowitz},
of the $\fj$ resonance.

CLEO II is a general purpose detector\cite{cleo} using
the $e^+ e^-$ storage ring, CESR\cite{cesr}, 
operating at $\sqrt{s}\sim \! 10.6$ GeV.  
CLEO II contains three concentric wire chambers
that detect charged particles over 95\% of the solid angle.
A superconducting solenoid provides a magnetic field of 1.5 T, giving
a momentum resolution of  
$\sigma_p/p \approx 0.5 \%$ for $p = 1$ GeV/$c$.  
Outside of the wire chambers and a time of flight system, but inside 
the solenoid, is a CsI electromagnetic calorimeter, 
consisting of 7800 crystals arranged
as two endcaps and a barrel region.  
For a 100 MeV electromagnetic shower in the barrel,
the calorimeter achieves an energy resolution of
$\sigma_E/E \approx 4\%$.

In two-photon events,
the initial state photons 
are approximately real and
tend to have a large fraction
of their momenta along
the beam line.
The electron and positron rarely 
have enough transverse momentum to be observed.
As the two photons generally have unequal momentum,
the
$\gam$ center of mass tends to be boosted along
the beam axis.  
We detect those events
in which the decay products have sufficient transverse momentum
to be observed in CLEO.  

We search for the two-photon production of $\fj$ in its
decay to $\ksks$
with each $K_s$ decaying into $\pppm$:

\begin{center}
\unitlength=1mm
\begin{picture}(22,15)(0,5)
        \put(2,16){\makebox(0,0){$\ggfkk$}}
        \put(32,10){\makebox(0,0){$\pptwo$}}
        \put(26,4){\makebox(0,0){$\ppone$}}
        \put(20,13){\line(0,-1){4}}
        \put(14,13){\line(0,-1){9}}
        \put(20,9){\vector(1,0){4}}
        \put(14,4){\vector(1,0){4}}
\end{picture}
\end{center}

\noindent In our analysis of $\lumi$ of data, 
we use the following selection criteria to minimize background.
We select events with four tracks.  We require that the sum of charges
is zero, the event energy is less than 6.0 GeV, and the transverse
compoment of the vector sum of the track momenta is less than
0.2 GeV/$c$.
To suppress $\gamma\gamma \rightarrow 4\pi$ we require 
two $\pi^+\pi^-$ pairs to form $K_s$ vertices 
separated in the $r-\phi$ plane by more than $5 \, {\rm mm}$.
Finally, we evaluate the $\pi^{\pm}$ track parameters at the
respective vertices, and select events in which 
$[m\ppone,m\pptwo]$ lies within a circle of radius 10 MeV
about the point $[m_{K_s},m_{K_s}]$.  
The detector
$K_s$ mass resolution is 3.3 MeV.

\begin{figure}[htbp]
   \centering \leavevmode
   \setfigsize
   \epsfbox{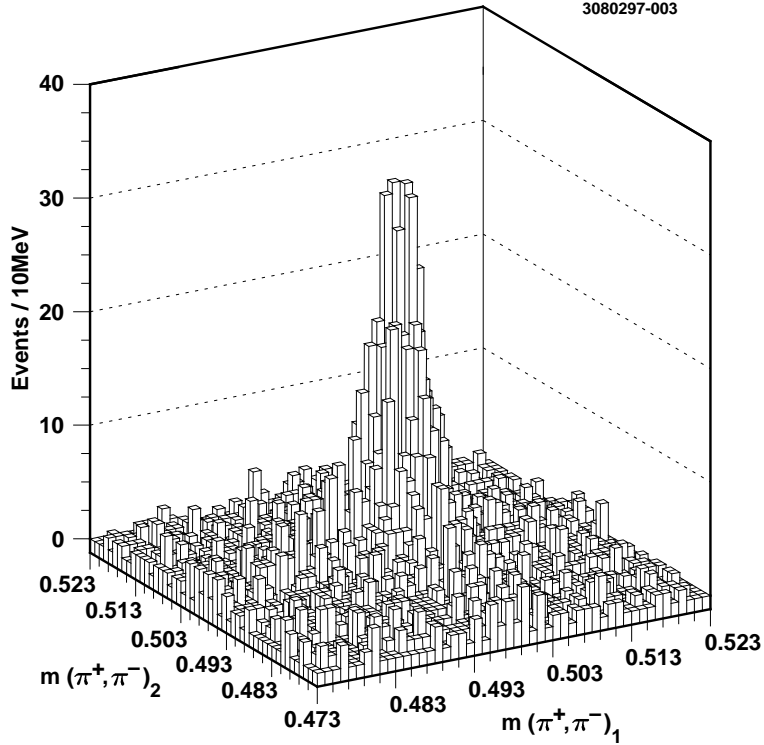}
   \caption{
       $m\ppone$ versus $m\pptwo$ for data.  Each event has
two entries corresponding to transposition of the labels 
$1\leftrightarrow 2$.}
   \label{fig_mpp_v_mpp}
\end{figure}

The distribution of $m\ppone$ versus $m\pptwo$ observed in
data is displayed in Figure \ref{fig_mpp_v_mpp} with all
selection criteria applied except the mass circle requirement.
There is a strong enhancement near the $[m_{K_s},m_{K_s}]$ point
in the $[m\ppone,m\pptwo]$ mass plane.  
After applying the 10 MeV mass circle criterion,
there is little non-$K_s$ background.

We use a Monte Carlo simulation to
determine our sensitivity to the
two-photon production of the $\fj$.
The two-photon Monte Carlo events were
generated using a program based on the BGMS formalism\cite{Budnev}.
For the simulation we assume the value $J=2$ for the 
total angular momentum.
We use a mass and width 
determined by 
combining\ifchklen\cite{besres}\else\footnote{\footbesres} \fi
the Mark III \cite{mark} and BES \cite{bes} 
results,
giving
$m_{f_J} = 2234\pm6 \, {\rm MeV}$ and 
$\Gamma_{f_J} = 19\pm11 \, {\rm MeV}$.
The simulation 
of the transport and decay
of the final state particles through the CLEO detector
is performed by a GEANT-based detector simulator\cite{geant}.
From the detector simulation we find a $\ksks$ mass
resolution, $\sigma_{K_sK_s}$, of $9 \, {\rm MeV}$
for $m_{K_sK_s}$ near 2.23 GeV.
The net selection efficiencies are 0.07 and 
0.15 for pure helicity 0 and pure helicity 2 respectively.

\begin{figure}[htbp]
   \centering \leavevmode
   \ifchklen
     \setfigsize
   \else
     \setbigfigsize
   \fi
   \epsfbox[93 356 515 686]{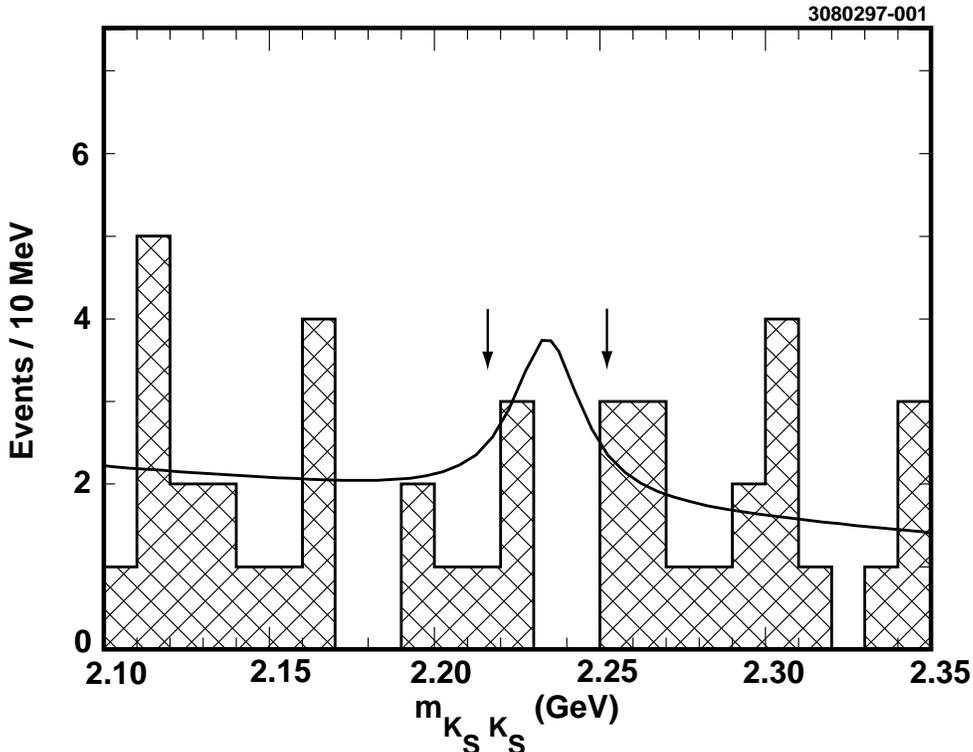}
   \caption{ $\ksks$ mass distribution (GeV) observed
in data near the $\fj$ mass. 
The vertical bars delineate the
signal region in which events are counted.  
The solid line is the sum of a fit to the 
background and the signal 
lineshape corresponding to the observed 95\% C.L. upper 
limit of 4.9 signal events.}

   \label{fig_m_ks_ks_fit}
\end{figure}

We construct a $\ksks$ mass distribution
for those events that satisfy all of the selection criteria.
In Figure \ref{fig_m_ks_ks_fit}, 
we display the data for the $\ksks$
mass region of interest.  
No enhancement at the $\fj$ mass is observed.

To determine the number of $\ggfj$ events, we count the
number of events within a region that has been 
optimized based on the lineshape of the $\fj$. 
In order to eliminate
dependence of the result on uncertainties in
the mass and width of the $\fj$, 
we construct nine limits, 
varying these resonance parameters by $\pm 1 \, \sigma$.
We convolve a detector 
resolution function with a Breit-Wigner resonance
to determine the expected shape.  
This lineshape is used to determine the signal region size 
that maximizes $\varepsilon^2/b$, where $\varepsilon$ is 
the fraction of the area under the signal lineshape that 
falls within the region, and $b$ is the estimated number of 
background events determined as described below.
For $\sigma_{K_sK_s} = 9 \, {\rm MeV}$ and 
$\Gamma_{f_J} = 19 \, {\rm MeV}$ this window is $\pm 18$ MeV,
for which $\varepsilon = 70\%$.  
For $\Gamma_{f_J} = 8$ and $30 \, {\rm MeV}$,
the window sizes are $\pm 13$ and $\pm 26$ MeV, respectively.

To obtain a background shape, 
we fit the $m_{\ksks}$ distribution with a linear function from 2.05 to 
2.35 GeV, excluding a $\pm 40$ MeV region centered on the 
expected mass.  From this we extract an average
background of 1.8$\pm 0.3$ events per 10 MeV at $m_{f_J}$ = 2.234 GeV.
Within the signal region determined for the central values
of the resonance parameters, we count four events.
Having observed four events while expecting 6.5 from background, 
we use the standard PDG technique of
extracting an upper limit for a Poisson
distribution with background \cite{pdg} 
to extract an upper limit of
$4.9$ signal events at the $95\% \, {\rm C.L.}$

To determine the value of $\brkkggfj$, we assume that
$\fj$ is produced incoherently with the background.
We scale the branching fraction and partial width used in
the Monte Carlo generator by the ratio of
the upper limit on the number of data events to the number of
selected Monte Carlo events, 
and by the ratio
of Monte Carlo to data luminosities,

\begin{eqnarray} 
  \Ggg^{data} \, \brdksks 
  &=& {n^{data} \over n^{MC} } \, { L^{MC} \over L^{data} } \,
  \Ggg^{MC} \, \brmcksks. 
  \label{eq_brgg_eq}
\end{eqnarray}

\noindent We repeat the entire analysis chain for nine
different sets of resonance parameters. 

The two-photon partial width, $\Ggg$, can be
expressed as the sum of two components, 
$\Ggg^{2,0}$ and $\Ggg^{2,2}$, 
the two-photon partial widths associated with helicity zero and
helicity two projections respectively.
We must differentiate between the two partial widths
because the detection efficiencies for the two allowed
helicity projections are not the same due
to their different final state angular distributions. 
Under the expectation that the ratio of  
$\Ggg^{2,2}$ : $\Ggg^{2,0}$ is 6:1 \cite{poppe,kolanoski} 
based on Clebsch-Gordon coefficients, 
we obtain the result,

\begin{eqnarray} 
  \brkkggfj
  &\leq& \glim, \ 95\% \, {\rm C.L.}
  \label{eq_brgg_val_6_1}
\end{eqnarray}

\noindent In Table \ref{limit_table} we present $\glim$ in eV
for $\pm 1 \, \sigma$ variation of the resonance mass and width.
The limits include uncertainties associated with systematics
which will be discussed later.

Without making any assumption about the ratio of 
partial widths of the two helicity projections,
we can set a 95\% C.L. functional limit,

\begin{eqnarray} 
  (0.52 \, \Ggg^{2,0} + 1.08 \, \Ggg^{2,2}) \, \brksks
  &\leq& \glim, \ 95\% \, {\rm C.L.}
  \label{eq_brgg_val}
\end{eqnarray}

\noindent The ratio of the partial width coefficients in
Equation \ref{eq_brgg_val} is
given by the ratio of efficiencies for helicity
zero to helicity two.
The overall normalization is set to be consistent with
Equation \ref{eq_brgg_val_6_1}.

Systematic uncertainties have been included in
determining these upper limits using a Monte Carlo program. 
We estimate the following systematic uncertainties in the
overall detector efficiency:
8\% due to triggering,
7\% due to tracking, and
7\% due to simulation of selection criteria.
The total systematic uncertainty associated with
efficiency is 13\%.
We estimate the systematic uncertainty in the 
background normalization to be 16\%.

\begin{table}[tbh]
\centering
\ifchklen
\else
\begin{minipage}{4in}
\fi
\begin{tabular}{cccccc}
& & & \multicolumn{3}{c}{\smash{Resonance Width, $\Gamma_{\fj}$}} \\
&     &\vline &8 MeV  &{\bf 19 MeV}         &30 MeV\\ 
\cline{2-6}
\smash{\lower 1.2ex \hbox{$m_{\fj}$}}
&2.230  &\vline &1.2 eV   &1.2 eV                   &1.3 eV \\ 
\smash{\lower 1.2ex \hbox{(GeV)}}
&{\bf 2.234} &\vline &1.2 eV   &{\bf 1.3 eV}             &1.5 eV \\ 
&2.238       &\vline &1.4 eV   &1.3 eV                   &1.8 eV \\ 
\end{tabular}
\caption{The upper limits, $\glim$ (eV), 
determined for $1 \sigma$ variations in
the resonance parameters.  
The central values are indicated in boldface.}
\label{limit_table}
\ifchklen
\else
\end{minipage}
\fi
\end{table}

We have verified our analysis by using the
same Monte Carlo simulation and analysis approach to measure
the two-photon partial width of the $\fpm$.
The $\fpm$ measurement is a sound test as the $\fpm$ produces
a prominent peak in the $K_s K_s$ mass distribution 
and has quantum numbers consistent with those
expected for the $\fj$.
We measure a value for the product of the 
partial width and the $K_sK_s$ branching fraction that is
within one standard deviation of the
PDG central value \cite{pdg} of $\brkkggfpm = 22\, {\rm eV}$.

The small value of the $\brkkggfj$ upper limit obtained
from this analysis supports the 
identification of the $\fj$ as a glueball.
We can make a more quantitative statement
by calculating the stickiness of the resonance.
Stickiness is a useful glueball figure of merit that
is a measure of color charge relative to electric charge. 
The definition of stickiness is \cite{chanowitz}: 

\begin{eqnarray} 
  S_{X}
  &\equiv&
  N_{l} {\left( {m_X \over k_{\psi \rightarrow \gamma X} } \right)}^{2l+1} \,
  { \Gamma (J/\psi \rightarrow \gamma X) \over 
    \Gamma (X \rightarrow \gamma \gamma) } \nonumber \\
  &\sim&
  { \left| \langle X| gg \rangle \right|^2 \over 
        \left| \langle X| \gam \rangle \right|^2 } 
  \label{eq_sticky_psuedo}
\end{eqnarray}

\noindent The parameter 
$k_{\psi \rightarrow \gamma X} = (m_{\psi}^2 - m_X^2)/(2m_{\psi})$
is the energy of the photon from a radiative $J/\psi$ decay
in the $J/\psi$ rest frame.
The phase-space term removes the mass dependence.
The quantum number $l$ indicates the 
angular momentum between the initial state gauge bosons.
$N_{l}$ is a normalization parameter defined so  
that the stickinesses of the $f_2(1270)$ $(l=0)$ is 1. 
To determine the value of $N_l$ we use the 
resonance mass, 
two-photon width, 
and radiative $J/\psi$ decay branching fraction 
given by the PDG \cite{pdg}.

To calculate a stickiness lower limit, 
we combine
our upper limit for $\brkkggfj = 1.3 \, {\rm eV}$
(evaluated at the central values of the resonance parameters)
with a 
value for $\gammajf$  
obtained by 
combining\ifchklen\cite{bespsiwidth}\else\footnote{\footbespsiwidth} \fi
results from Mark III \cite{mark} and BES\cite{bes}.
The 
${\cal B}(J/\psi \rightarrow \gamma \fj) \, {\cal B}(\fj \rightarrow K_s K_s)$ 
branching fraction so
determined 
is $(2.2 \pm 0.6) \times 10^{-5}$.
From this we calculate
a lower limit on stickiness of 82 at the 95\% C.L.
for the $\fj$.  
The statistical and systematic uncertainties of the inputs,
including the uncertainty on the $J/\psi$ branching fraction, 
are incorporated into this limit through a Monte Carlo program.  
This lower limit is much larger than the
value of one expected for a $\qqb$ resonance.

The observation of the $\fj$'s in ``glue rich'' environments
such as the radiative $J/\psi$ decay has made it a
glueball candidate.
With the limit on $\brkkggfj$ presented
here we are able to make a much stronger statement.
In particular,
it is difficult to explain how a $\qqb$ meson,
even pure $s \overline{s}$,
could have such a large stickiness.
In general, explanations that give small
two-photon partial widths give small
radiative $J/\psi$ decay branching fractions.
Radial and angular excitations fall into this category.
A $J = 4$ resonance is not ruled out experimentally.
However, under the assumption $J=4$, 
the phase space term to which stickiness is proportional
becomes very large.  
A small two photon width due to a cancelation involving
specific values of the singlet-octet mixing and the
ratio of matrix elements is possible but unlikely.

In this Letter we have presented the results of 
the search for $\fj$ production in two-photon interactions.
We have reported a very small upper limit for $\brkkggfj$.
The minimum stickiness obtained from 
the two-photon width upper limit
is difficult to understand
in the context of a $\qqb$ resonance,
and should be considered as strong evidence that the
$\fj$ is a glueball.

We gratefully acknowledge the effort of the CESR staff in providing us with
excellent luminosity and running conditions.
We would like to thank 
M. Chanowitz for his thoughtful comments.
This work was supported by 
the National Science Foundation, 
the U.S. Department of Energy, 
the Heisenberg Foundation,
the Alexander von Humboldt Stiftung,
the Natural Sciences and Engineering Research Council of Canada,
and the A. P. Sloan Foundation. 



\begin{thebibliography}{99}
   \bibitem{mark}
      Mark III Collaboration, R. Baltrusaitis {\sl et al.},
      {\it Phys. Rev. Lett.}
      {\bf 56}, 107 (1986).
   \bibitem{bes}
      BES Collaboration, J.Z. Bai {\sl et al.},
      {\it Phys. Rev. Lett.}
      {\bf 76}, 3502 (1996).
   \bibitem{gams}
      GAMS Collaboration, Alde {\sl et al.},
      {\it Phys. Lett. B}
      {\bf 177}, 120 (1986).
   \bibitem{lass}
      LASS Collaboration, D. Aston {\sl et al.},
      {\it Phys. Lett. B}
      {\bf 215}, 199 (1988).
   \bibitem{mss}
      MSS Collaboration, B.V. Bolonkin {\sl et al.},
      {\it Nuc. Phys. B}
      {\bf 309}, 426 (1988).
   \bibitem{michael}
      C. Michael
      ``Glueballs and Hybrid Mesons''
      hep-ph/9605243 (1996).   
   \bibitem{morningstar}
      C. Morningstar and M. Peardon,
      {\it Nucl. Phys. Proc. Suppl. }
      {\bf 53}, 917 (1997),      
      hep-lat/9608050.
   \bibitem{argus}
      ARGUS Collaboration, H. Albrecht {\sl et al.},
      {\it Z. Phys. C}
      {\bf 48}, 183 (1990).
   \bibitem{chanowitz}
      M. Chanowitz
      ``Resonances in Photon-Photon Scattering''
      Proceedings of the VI$^{th}$ International Workshop on Photon-
      Photon Collisions (1984). 
   \bibitem{cleo}
      CLEO Collaboration, Y. Kubota {\sl et al.},
      {\it Nucl. Inst. \& Meth.}
      {\bf A320}, 66 (1992).
   \bibitem{cesr}
      D. Rubin
      {\it Proceedings of the 1995 Particle Accelerator Conference} 
      {\bf 1}, 481 (1995).
   \bibitem{Budnev}
      V.M.Budnev, I.F.Ginzburg, G.V.Meledin, and V.G.Serbo, 
      {\it Phys. Rep.} {\bf 15C}, 181 (1975).
   \ifchklen
   \bibitem{besres}
     \footbesres
   \fi
   \bibitem{geant}
     R.Brun {\sl et al.},
     ``GEANT3 Users Guide,''
     CERN DD/EE/84-1 (1987).
   \bibitem{pdg}
      Particle Data Group,
      {\it Phys. Rev. D}
      {\bf 54} (1996).
    \bibitem{poppe}
      M. Poppe,
      {\it Int. Jour. Mod. Phys.} {\bf A1} 545 (1986).
   \bibitem{kolanoski}
      H. Kolanoski and P. Zerwas,
      ``Two-Photon Physics''
      {\it High Energy Electron-Positron Physics }, 
      World Scientific, Singapore (1988). 
   \ifchklen
   \bibitem{bespsiwidth}
     \footbespsiwidth
   \fi

\end{thebibliography}
\end{document}